\documentclass[prd,preprint,showpacs]{revtex4}
\usepackage{graphicx}
\begin{document}
\title{Probing the curvature and dark energy}
\author{Yungui Gong}
\affiliation{College of Electronic Engineering, Chongqing
University of Posts and Telecommunications, Chongqing 400065,
China}
\email{gongyg@cqupt.edu.cn}
\author{Yuan-Zhong Zhang}
\affiliation{ CCAST (World Laboratory), P.O. Box 8730, Beijing 100080, China \\
Institute of Theoretical Physics, Chinese Academy of Sciences,
P.O. Box 2735, Beijing 100080, China}
\begin{abstract}
Two new one-parameter tracking behavior dark energy representations
$\omega=\omega_0/(1+z)$ and $\omega=\omega_0
e^{z/(1+z)}/(1+z)$ are used to probe the geometry of the Universe
and the property of dark energy. The combined type Ia supernova
(SN Ia), Sloan Digital Sky Survey (SDSS), and Wilkinson Microwave
Anisotropy Probe (WMAP) data indicate that the Universe is almost
spatially flat and that dark energy contributes about 72\% of the
matter content of the present universe. The observational data
also tell us that $\omega(0)\sim -1$. It is argued that the
current observational data can hardly distinguish different
dark energy models to the zeroth order. The
transition redshift when the expansion of the Universe changed from deceleration phase to acceleration
phase is around $z_{\rm T}\sim 0.6$ by using our
one-parameter dark energy models.
\end{abstract}
\preprint{astro-ph/0502262}
\maketitle

\section{Introduction}

The SN Ia data suggest that the Universe is dominated by dark
energy \cite{sp99,gpm98,agr98}. Since 1998, many dark energy
models have been proposed in the literature. The simplest dark
energy model is the cosmological constant model. However, the
smallness of the value of the observed cosmological constant has
puzzled theoretical physicists for a long time. For a review of
dark energy models, see, for example references \cite{review} and
\cite{padmanabhan03}. Although there exist a lot of dark energy
models, we are still not able to decide which model gives us the
right answer and find out the nature of dark energy. From
theoretical point of view, perhaps the lack of understanding of
quantum gravity is the main reason. To advance our understanding
of dark energy, we may use observational data to probe the nature
of dark energy. It is not practical to test every single dark
energy model by using the observational data. Therefore, a model
independent probe of dark energy is one of the best choices to
study the nature of dark energy.

The usual model independent method is through parameterizing dark
energy or the equation of state parameter $\omega(z)$ of dark
energy. The simplest method is parameterizing $\omega(z)$ as a
constant. To model the dynamical evolution of dark energy, we can
parameterize $\omega(z)$ as the power law expansion
$\omega(z)=\sum_{i=0}^N \omega_i z^i$ \cite{jwaa,hut,jwaa1,pastier}.
Recently, a simple two-parameter model
$\omega(z)=\omega_0+\omega_a z/(1+z)$ was extensively discussed
\cite{polarski,linder,choudhury,feng}. Jassal, Bagla and
Padmanabhan later modified this two-parameter model as
$\omega(z)=\omega_0+\omega_a z/(1+z)^2$ \cite{hkjbp}. More
complicated forms of $\omega(z)$ were also discussed in the
literature \cite{gong04,efstathiou,gefstathiou,pscejc,wetterich}.
Instead of parameterizing $\omega(z)$, we can also parameterize
the dark energy density itself, like a simple power law expansion
$\Omega(z)=\sum_{i=0}^N A_i z^i$
\cite{alam,alam1,daly,daly1,gong,jonsson} and the piecewise
constant parameterization \cite{wang,wang1,cardone,huterer}. In
\cite{gong04}, Gong used the SN Ia data to discuss some
two-parameter representations of dark energy in a spatially flat
cosmology. It was found that the SN Ia data marginally favored a
phantom-like dark energy model. It was also found that the
transition redshift $z_{\rm T}\sim 0.3$. In this paper, we propose
two one-parameter dark energy models $\omega=\omega_0/(1+z)$ and
$\omega=\omega_0 e^{z/(1+z)}/(1+z)$ and we use the SN Ia, the SDSS
and the WMAP data to probe the geometry of the Universe. We also
compare these two models with two two-parameter dark energy
models.

This paper is organized as follows. In section II, We first use
the $\Lambda$CDM model as an example to show the method of
fitting the whole 157 gold sample of SN Ia data compiled in
\cite{riess}, the parameter $A$ measured from the SDSS data
\cite{sdss} and the shift parameter $\mathcal{R}$ measured from
the WMAP data \cite{wang,wang1} to a dark energy model. The parameter $A$ is a dark energy model
independent parameter found by Eisenstein et al. in \cite{sdss} when they analyzed the large scale
correlation function of a large spectroscopic sample of luminous, red galaxies for SDSS.
It is related to the path from $z=0$ to $z=0.35$. $\mathcal{R}$ is the
shift of the positions of the acoustic peaks
in the angular power spectrum due to the effect of changing the values
of $\Omega_{m0}$ and $\Omega_{k0}$ on the Cosmic Microwave Background anisotropy. In section
III, we propose two new tracking behavior one-parameter dark
energy representations $\omega=\omega_0/(1+z)$ and
$\omega=\omega_0 e^{z/(1+z)}/(1+z)$. In section IV, we fit the models to the
observational data. In section V, we fit two two-parameter
parameterizations $\omega=\omega_0+\omega_a z/(1+z)$ and
$\omega=\omega_0+\omega_a z/(1+z)^2$ to the observational data. In
section VI, we conclude the paper with some discussion.

\section{$\Lambda$CDM model with curvature} For the simplest
$\Lambda$CDM model where the dark energy is the cosmological
constant, i.e., $\rho=-p=\Lambda$, we have \begin{equation}
\label{cdmcos}
H^2=H^2_0[\Omega_{m0}(1+z)^3+\Omega_{r0}(1+z)^4-\Omega_{k0}(1+z)^2+1+\Omega_{k0}-\Omega_{m0}-\Omega_{r0}],
\end{equation} where $\Omega_m\,(\Omega_r)=8\pi
G\rho_m\,(\rho_r)/3H_0^2$, $\Omega_{r0}=8.35\times 10^{-5}$
\cite{wmap}, and $\Omega_k=k/a^2 H^2_0$. The parameters
$\Omega_{m0}$ and $\Omega_{k0}$ are determined by minimizing
\begin{equation}\label{chi2} \chi^2=\sum_i{[\mu_{\rm
obs}(z_i)-\mu(z_i)]^2\over
\sigma^2_i}+\frac{(A-0.469)^2}{0.017^2}+\frac{(\mathcal{R}-1.716)^2}{0.062^2},\end{equation}
where the extinction-corrected distance modulus
$\mu(z)=5\log_{10}(d_{\rm L}(z)/{\rm Mpc})+25$, the luminosity
distance is \begin{eqnarray} \label{lumdis} d_{\rm L}(z)
=a_0(1+z)r(z)&=&\frac{a_0(1+z)}{\sqrt{|k|}}{\rm sinn}
\left[\frac{\sqrt{|k|}}{a_0H_0}\int_0^z
\frac{dz'}{E(z')}\right]\nonumber\\
&=&\frac{1+z}{H_0\sqrt{|\Omega_{k0}|}} {\rm
sinn}\left[\sqrt{|\Omega_{k0}|}\int_0^z
\frac{dz'}{E(z')}\right],
\end{eqnarray}
${\rm sinn}(\sqrt{|k|}x)/\sqrt{|k|}=\sin(x)$, $x$, $\sinh(x)$ if $k=1$, 0, $-1$,
the dimensionless Hubble parameter $E(z)=H(z)/H_0$, the parameter
$A$ is defined as \cite{sdss}
\begin{equation}
\label{para}
A=\frac{\sqrt{\Omega_{m0}}}{0.35}\left[\frac{0.35}{E(0.35)}\frac{1}{|\Omega_{k0}|}{\rm
sinn}^2\left(\sqrt{|\Omega_{k0}|}\int_0^{0.35}
\frac{dz}{E(z)}\right)\right]^{1/3}=0.469\pm 0.017,
\end{equation}
the shift parameter \cite{wang,wang1}
\begin{equation}
\label{shift}
\mathcal{R}=\frac{\sqrt{\Omega_{m0}}}{\sqrt{|\Omega_{k0}|}}{\rm
sinn}\left(\sqrt{|\Omega_{k0}|}\int_0^{z_{ls}}\frac{dz}{E(z)}\right)=1.716\pm
0.062,
\end{equation}
$z_{ls}=1089\pm 1$ \cite{wmap} and $\sigma_i$ is the total
uncertainty in the SN Ia data. In other words, we use the 157 gold
sample SN Ia data compiled in \cite{riess}, the parameter $A$
measured from the SDSS data \cite{sdss} and the shift parameter
$\mathcal{R}$ measured from the WMAP data \cite{wang,wang1} to
find out the parameters $\Omega_{m0}$ and $\Omega_{k0}$. The
nuisance parameter $H_0$ appeared in Eq. (\ref{lumdis}) is
marginalized over with a flat prior assumption. Since $H_0$
appears linearly in the form of $5\log_{10}H_0$ in $\chi^2$, so
the marginalization by integrating $\mathcal{L}=\exp(-\chi^2/2)$ over all
possible values of $H_0$ is equivalent to finding the value of
$H_0$ which minimizes $\chi^2$ if we also include the suitable
integration constant and measure function.

\begin{figure}
\centering
\includegraphics[width=12cm]{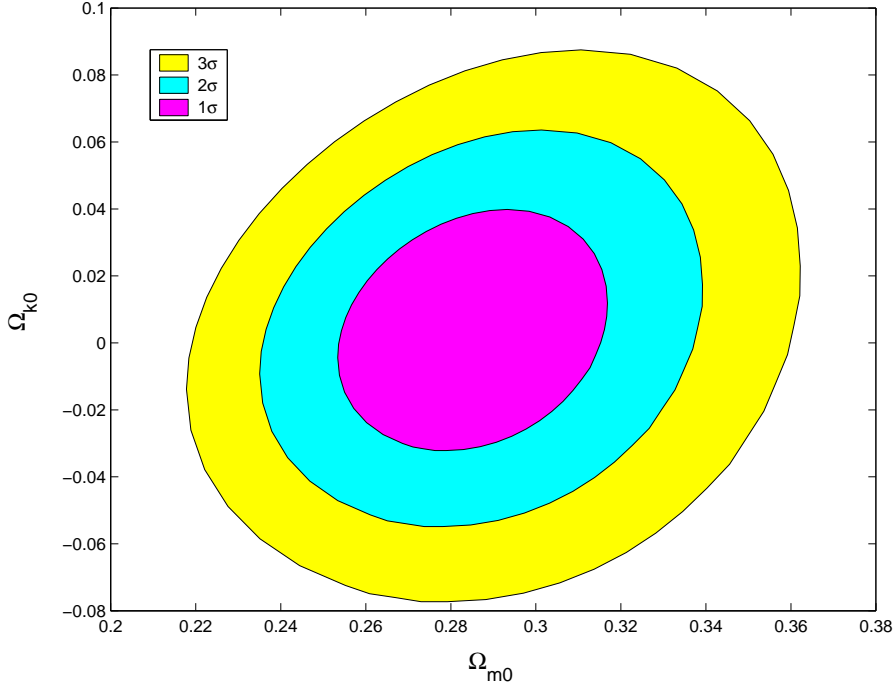}
\caption{The $1\sigma$, $2\sigma$ and $3\sigma$ contour plots of $\Omega_{m0}$ and $\Omega_{k0}$ for
the model with the cosmological constant as dark energy}
\label{cdmcont}
\end{figure}

The best fit parameters to the combined SN Ia, SDSS and WMAP data
are $\Omega_{m0}=0.28\pm 0.03$ and $\Omega_{k0}=0.004\pm 0.04$
with $\chi^2=177.14$. The contour plot of $\Omega_{m0}$ and
$\Omega_{k0}$ is shown in Fig. \ref{cdmcont}.

\section{One-parameter parameterization}
The main goal of this work is to study the geometry of the Universe and the property
of dark energy by using observational data. In the Introduction, we mentioned
several different parameterizations. Those parameterizations have two
or more parameters, so it is difficult to use those parameterizations to investigate the geometry of the Universe
because we have to add two more cosmological parameters $\Omega_{m0}$ and $\Omega_{k0}$ to the model. Therefore, it
is better that the parameterization has only
one parameter so that the whole model has three cosmological parameters. Of course, this
kind of parameterization assumed that the dark energy has evolution and it is
not a cosmological constant.

Model 1: To make the parameterization physical, we look for tracking behavior representation, i.e.,
we require $\omega(z=\infty)=0$.
We first consider a simple one-parameter dark
energy representation
\begin{equation}
\label{1pna} \omega(z)=\frac{\omega_0}{1+z}.
\end{equation}
At early times, $z\gg 1$, $\omega(z)\sim 0$. In the far
future, $1+z\rightarrow 0$ and $\omega(z)\rightarrow -\infty$.
This simple parameterization has future singularity. The energy conservation equation of the dark energy can be
written as
\begin{equation}
\label{conserv} \frac{d\ln\rho}{dz}=\frac{3(1+\omega)}{1+z}.
\end{equation}
The acceleration equation is
\begin{equation}
\label{acceq}
\frac{\ddot{a}}{a}=-\frac{4\pi G}{3}(\rho_m+2\rho_r+\rho+3p).
\end{equation}
Combining Eqs. (\ref{1pna}) and (\ref{conserv}), we get
\begin{equation}
\label{dens1pa} \rho=\rho_0(1+z)^3\exp\left(\frac{3\omega_0
z}{1+z}\right).
\end{equation}
It is obvious that $\rho\sim e^{3\omega_0}\rho_0(1+z)^3$ when
$z\gg 1$ and $\rho\rightarrow \infty$ when $z\rightarrow -1$.
At early times, the energy density looks like matter with
effective $\Omega_{m0}=e^{3\omega_0}\Omega_0$. Here $\Omega=8\pi
G\rho/3H_0^2$ and
$\Omega_0=1+\Omega_{k0}-\Omega_{m0}-\Omega_{r0}$. This model may
be thought as a unified model of dark matter and dark energy. The
sound speed of the model is
\begin{equation}
\label{cs1pa}
c^2_s=\frac{\partial p}{\partial
\rho}=\frac{2\omega_0(1+z)+3\omega^2_0}{3(1+z)^2+3\omega_0(1+z)}.
\end{equation}
So $c^2_{s0}=(2\omega_0+3\omega^2_0)/(3+3\omega_0)\neq 0$. When $1+z+\omega_0>0$ and $1+z+1.5\omega_0<0$,
$c^2_{s0}>0$. For any unified theory of dark energy and dark
matter, it was shown that
the wave number $k$ dependence of density perturbation growth due to the presence of a nonzero sound speed
for a period of time
produces unphysical oscillations or exponential blow-up in the
matter power spectrum \cite{sandvik,bento04,rrries}. Therefore this model as
the unified model of dark matter and dark energy
is not feasible.  The model
with interactions between dark energy and dark matter was
discussed in \cite{guo}.

Now we consider the model as a dark energy model. As
we saw above, the dark energy behaves as ordinary matter at early times,
we can interpret this as the tracking behavior, i.e.,
the dark energy tracked the matter at early times. The total
effective matter density is $\Omega_{m0}^{\rm
eff}=\Omega_{m0}+e^{3\omega_0}\Omega_0$, and we expect that
$e^{3\omega_0}\Omega_0\ll \Omega_{m0}$. Substitute Eqs.
(\ref{1pna}) and (\ref{dens1pa}) into Eq. (\ref{acceq}) and
neglect the radiation contribution, we find that the transition
redshift $z_{\rm T}$ satisfies the following equation
\begin{equation}
\label{zt1pna}
(1+z)^3\left[\Omega_{m0}+\left(1+\frac{3\omega_0}{1+z}\right)\Omega_0\exp\left(\frac{3\omega_0
z}{1+z}\right)\right]=0.
\end{equation}

Model 2: Now we consider another one parameter dark energy
parameterization
\begin{equation}
\label{1pnb} \omega(z)=\frac{\omega_0}{1+z}e^{z/(1+z)}.
\end{equation}
For this model, $\omega(z)\sim 0$ when $z\gg 1$. The major
difference between this model (\ref{1pnb}) and the model
(\ref{1pna}) is that $\omega(z)\rightarrow 0$ as $z\rightarrow -1$
for this model. These two models have almost the same behavior in the past and
very different behavior in the future. Combining Eqs. (\ref{conserv}) and (\ref{1pnb}),
we get
\begin{equation}
\label{dens1pb} \rho=\rho_0(1+z)^3\exp\left(3\omega_0
e^{z/(1+z)}-3\omega_0\right).
\end{equation}
It is obvious that $\rho\sim
e^{3\omega_0e-3\omega_0}\rho_0(1+z)^3$ when $z\gg 1$ and $\rho\sim
e^{-3\omega_0}\rho_0(1+z)^3$ when $z\rightarrow -1$. At early times, the energy density looks like matter with effective
$\Omega_{m0}=e^{3\omega_0e-3\omega_0}\Omega_0$, and it behaves
like matter with effective $\Omega_{m0}=e^{-3\omega_0}\Omega_0$ in
the far future too. This model may also be thought as a unified
model of dark matter and dark energy. The sound speed of the model
is
\begin{equation}
\label{cs1pb} c^2_s=\frac{\partial p}{\partial
\rho}=\frac{2\omega_0(1+z)e^{z/(1+z)}+\omega_0(1+3\omega_0e^{z/(1+z)})e^{z/(1+z)}}{3(1+z)^2+3\omega_0e^{z/(1+z)}}.
\end{equation}
So $c^2_{s0}=\omega_0< 0$ and it also produces exponential blow-up in the matter power spectrum.
This model as the unified model of dark matter and dark energy is not feasible too.
Again we consider this model as a dark energy
model. One key feature of this model is that the model behaves
like matter both in the past and the future. The Universe will
expand with deceleration in the future. In the past, the dark
energy tracked the matter. The total effective matter density is
$\Omega_{m0}^{\rm eff}=\Omega_{m0}+e^{3\omega_0(e-1)}\Omega_0$,
and we expect that $e^{3\omega_0(e-1)}\Omega_0\ll \Omega_{m0}$.
Substitute Eqs. (\ref{1pnb}) and (\ref{dens1pb}) into Eq.
(\ref{acceq}) and neglect the radiation contribution, we find that
the transition redshift $z_{\rm T}$ satisfies the following
equation
\begin{equation}
\label{zt1pnb}
(1+z)^3\left[\Omega_{m0}+\left(1+\frac{3\omega_0}{1+z}\exp(\frac{z}{1+z})\right)\Omega_0\exp\left(3\omega_0(e^{
z/(1+z)}-1)\right)\right]=0.
\end{equation}

\section{Data Fitting Results}
\begin{figure}
\centering
\includegraphics[width=12cm]{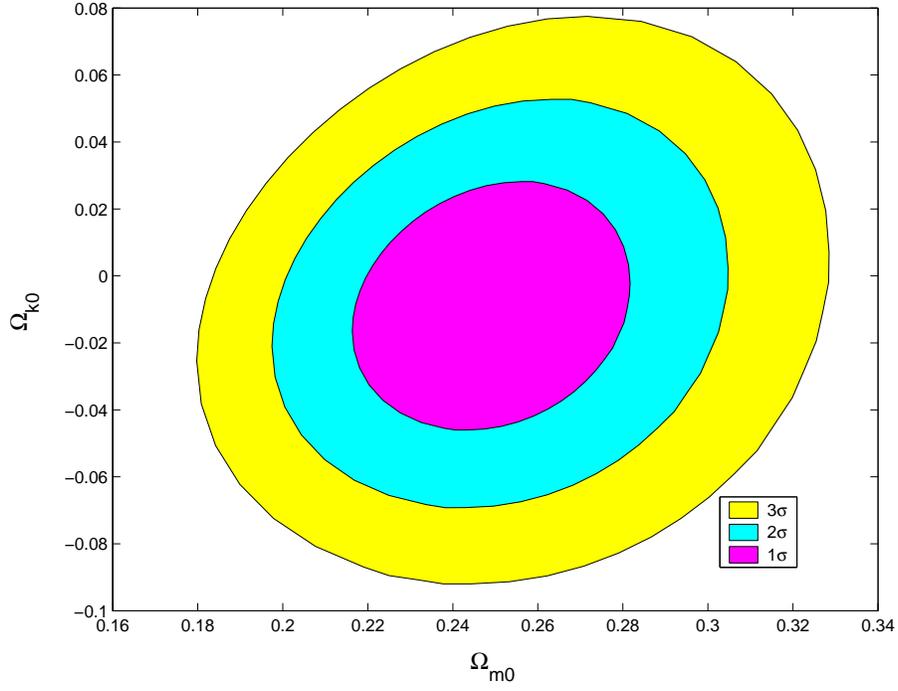}
\caption{The $1\sigma$, $2\sigma$ and $3\sigma$ contour plots of $\Omega_{m0}$ and $\Omega_{k0}$ for
the parameterization $\omega=\omega_0/(1+z)$.}
\label{cont1pna}
\end{figure}

\begin{figure}
\centering
\includegraphics[width=12cm]{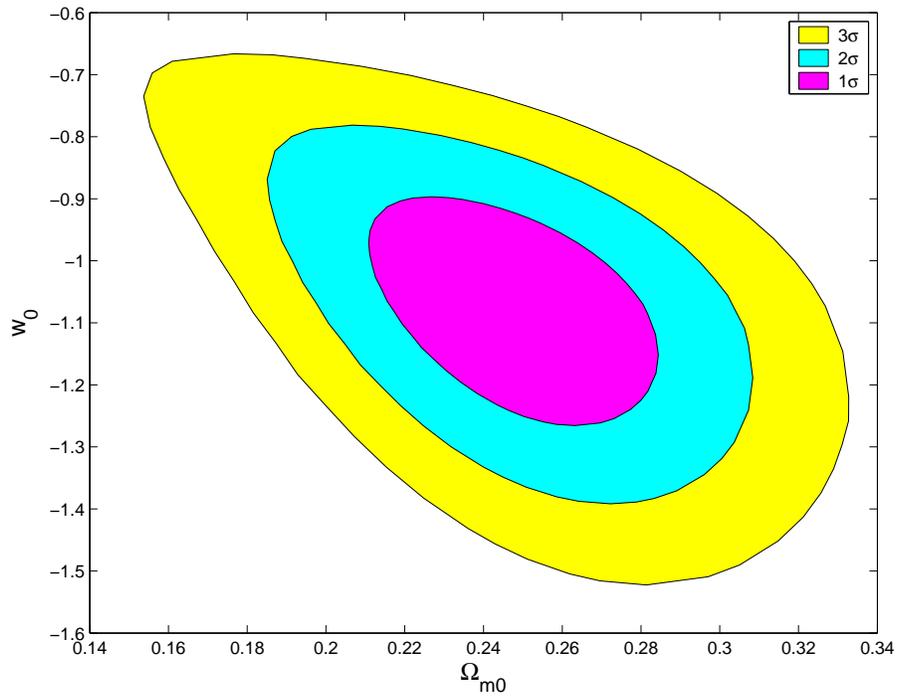}
\caption{The $1\sigma$, $2\sigma$ and $3\sigma$ contour plots of $\Omega_{m0}$ and $\omega_0$ for
the parameterization $\omega=\omega_0/(1+z)$.}
\label{cont1pna1}
\end{figure}

By fitting the model 1 to the combined SN Ia, SDSS and WMAP data,
we get $\Omega_{m0}=0.25\pm 0.05$, $\Omega_{k0}=-0.009\pm 0.05$
and $\omega_0=-1.1\pm 0.2$ with $\chi^2=175.4$. If we take the model 1
as a unified model of dark energy and dark
matter, we find that the best fit parameters to the combined SN
Ia, SDSS and WMAP data are $\Omega_{k0}=-0.05\pm 0.04$ and
$\omega_0=-0.42\pm 0.04$ with $\chi^2=203.6$. Since $\Delta\chi^2=203.6-175.4=28.2$, we
conclude that this model as a
unified model of dark matter and dark energy is not a
viable model. The contour plots
of $\Omega_{m0}$ and $\Omega_{k0}$ by fixing $\omega_0$ at its
best fit value $-1.1$ are shown in Fig. \ref{cont1pna}. The contour plots
of $\Omega_{m0}$ and $\omega_0$ by fixing $\Omega_{k0}$ at its
best fit value $-0.009$ are shown in Fig. \ref{cont1pna1}. The
evolution of $\omega(z)$ is shown in Fig. \ref{obswqz}. Substitute
the best fit values to Eq. (\ref{zt1pna}), we get the transition
redshift $z_{\rm T}=0.56$. The results are summarized in Table \ref{fittab}.

If we use SN Ia only, then we get
$\Omega_{m0}=0.33^{+0.3}_{-0.22}$,
$\Omega_{k0}=-0.33^{+1.19}_{-0.32}$ and $\omega_0=-7.5^{+6.9}_{-33.8}$
with $\chi^2=171.9$.

\begin{figure}
\centering
\includegraphics[width=12cm]{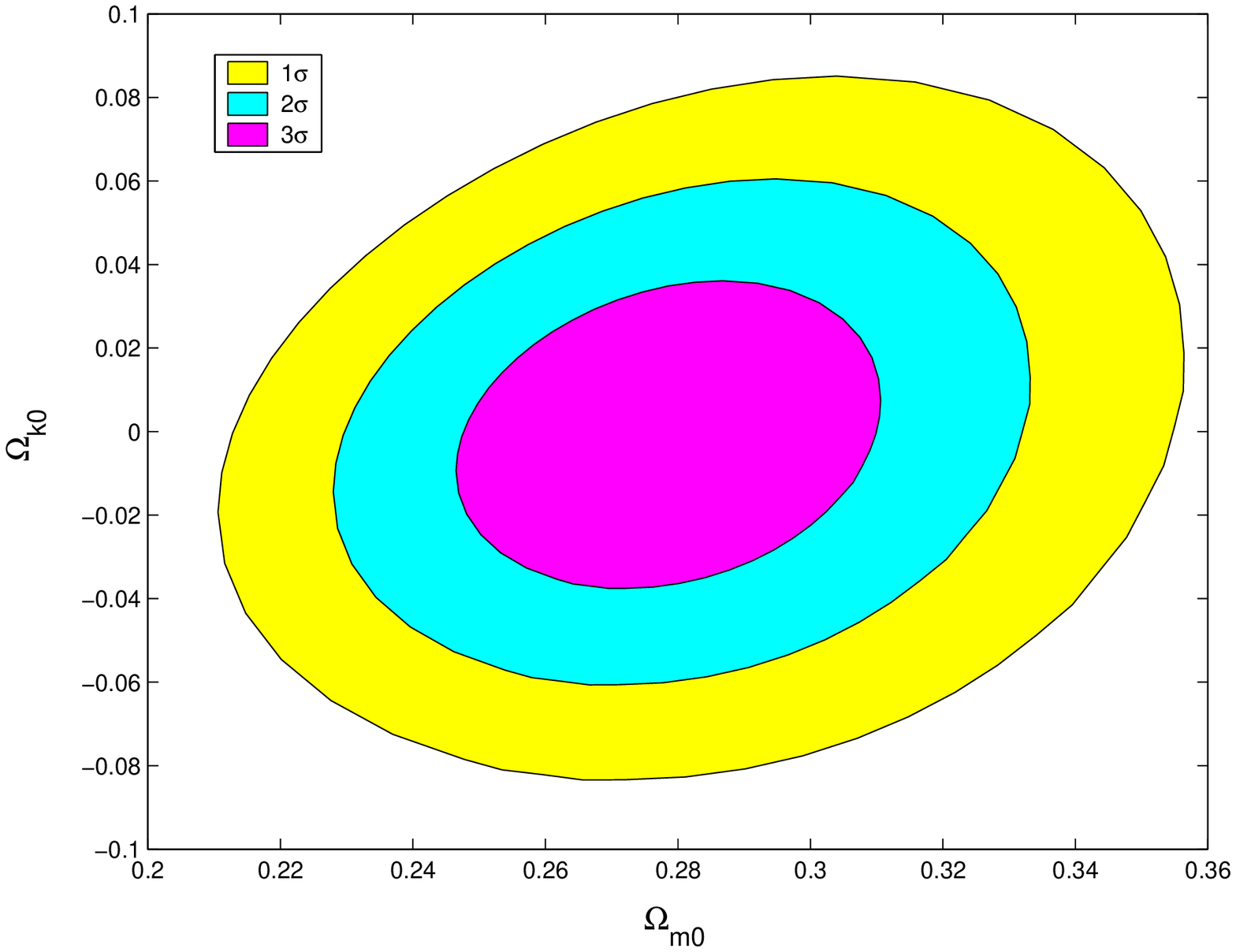}
\caption{The $1\sigma$, $2\sigma$ and $3\sigma$ contour plots of $\Omega_{m0}$ and $\Omega_{k0}$ for
the parameterization $\omega=\omega_0
\exp(z/(1+z))/(1+z)$.} \label{cont1pnb}
\end{figure}

\begin{figure}
\centering
\includegraphics[width=12cm]{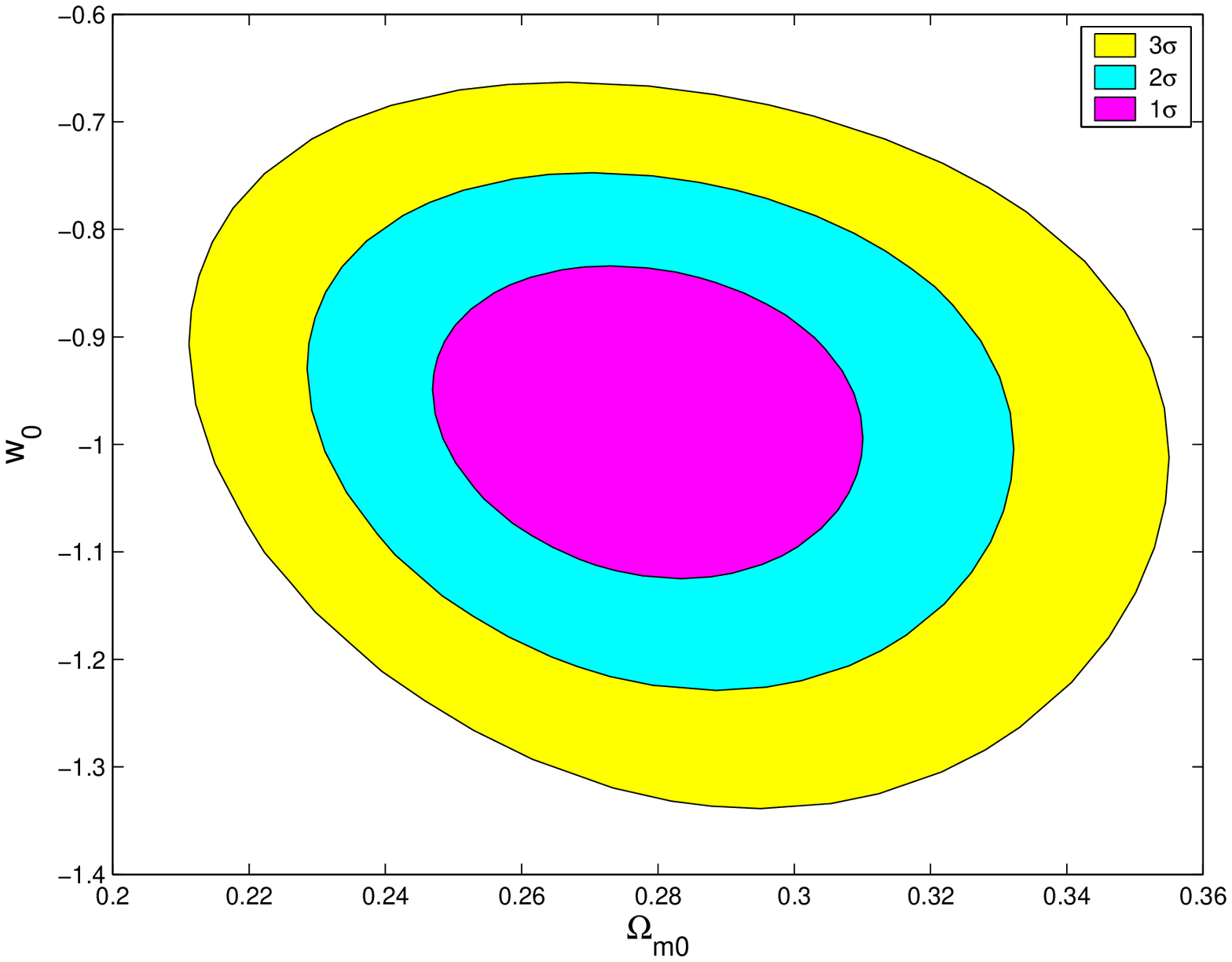}
\caption{The $1\sigma$, $2\sigma$ and $3\sigma$ contour plots of $\Omega_{m0}$ and $\omega_0$ for
the parameterization $\omega=\omega_0
\exp(z/(1+z))/(1+z)$.} \label{cont1pnb1}
\end{figure}

By fitting the model 2 to the combined SN Ia, SDSS and WMAP data,
we get $\Omega_{m0}=0.28\pm 0.04$,
$\Omega_{k0}=-0.001^{+0.046}_{-0.045}$ and
$\omega_0=-0.97^{+0.17}_{-0.19}$ with $\chi^2=176.5$. If we take the model 2
as a unified model of dark energy and dark
matter, the best fit parameters to the combined SN Ia, SDSS and WMAP data
are $\Omega_{k0}=-0.10\pm 0.05$ and
$\omega_0=-0.22^{+0.02}_{-0.03}$ with $\chi^2=233.2$. Again this
model as a unified model of dark matter and dark energy can be
firmly ruled out. The contour
plots of $\Omega_{m0}$ and $\Omega_{k0}$ by fixing $\omega_0$ at
its best fit value $-0.97$ are shown in Fig. \ref{cont1pnb}. The contour
plots of $\Omega_{m0}$ and $\omega_0$ by fixing $\Omega_{k0}$ at
its best fit value $-0.001$ are shown in Fig. \ref{cont1pnb1}. The
evolution of $\omega(z)$ is shown in Fig. \ref{obswqz}. Substitute
the best fit values to Eq. (\ref{zt1pnb}), we get the transition
redshift $z_{\rm T}=0.66$. The results are also shown in Table \ref{fittab}.

If we use SN Ia only, then we get $\Omega_{m0}=0.33_{-0.20}$,
$\Omega_{k0}=-0.34^{+1.19}_{-0.32}$ and
$\omega_0=-7.4^{+6.8}_{-33.4}$ with $\chi^2=171.9$.

Although the two models discussed in the previous section have
very different future behavior, they both fit the current data as
well as the $\Lambda$CDM model. This may suggest that the current
data fitting method cannot distinguish models with very different
future behavior. The best fit results also show that the Universe
is almost spatially flat, and that the best fit results using the
combined SN Ia, SDSS and WMAP data are different from those using
SN Ia alone.

\section{Two-parameter Parameterization}
In this section, we consider spatially flat cosmology only.

\begin{figure}
\centering
\includegraphics[width=12cm]{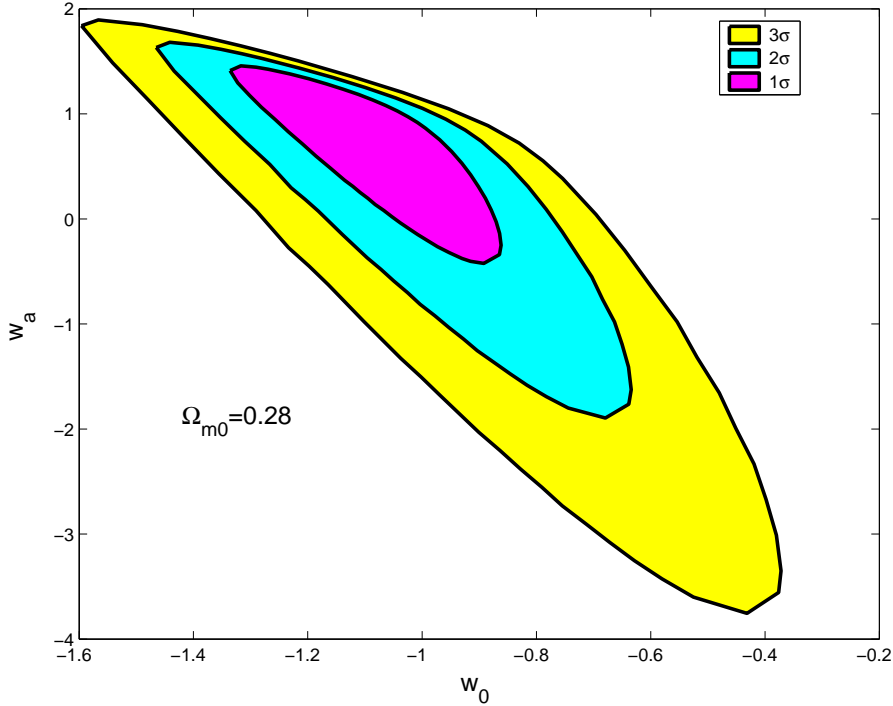}
\caption{The contour plot of $\omega_0$ and $\omega_a$ for the
parameterization $\omega=\omega_0+\omega_a z/(1+z)$} \label{wqzl}
\end{figure}

\begin{figure}
\centering
\includegraphics[width=12cm]{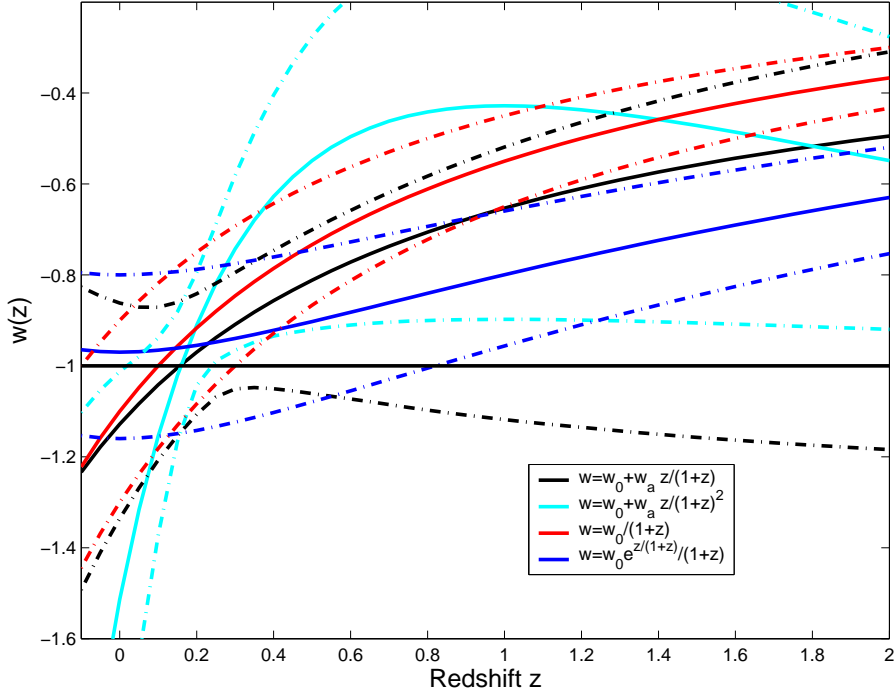}
\caption{The behavior of $\omega(z)$. The solid lines plot
$\omega(z)$ by using the best fit parameters and the dash dotted
lines are for 1$\sigma$ errors. For the two-parameter representations, the $1\sigma$
errors are obtained from the $1\sigma$ contours of $\omega_0$ and $\omega_a$.} \label{obswqz}
\end{figure}

Model 3: We first consider the parameterization \cite{polarski,linder}
\begin{equation}
\label{linder} \omega=\omega_0+{\omega_a z\over
1+z}.
\end{equation}
When $z\gg 1$, we have $\omega \sim \omega_0+\omega_a$.
$\omega\rightarrow \pm \infty$ when $z\rightarrow -1$. Combining
Eqs. (\ref{conserv}) and (\ref{linder}), we get the dark energy
density
\begin{equation}
\label{linddark}
\Omega=\Omega_0(1+z)^{3(1+\omega_0+\omega_a)}\exp(-3\omega_az/(1+z)),
\end{equation}
where $\Omega_0=1-\Omega_{m0}-\Omega_{r0}$. Substitute Eqs.
(\ref{linder}) and (\ref{linddark}) into Eq. (\ref{acceq}) and
neglect the radiation contribution, we find that $z_{\rm T}$
satisfies the following equation
\begin{equation}
\label{ztlind}
\Omega_{m0}+(1-\Omega_{m0})\left(1+3\omega_0+{3\omega_a z\over
1+z}\right)(1+z)^{3(\omega_0+\omega_a)}
\exp\left({-3\omega_az\over 1+z}\right)=0.
\end{equation}
The best fit to the combined SN Ia, SDSS and WMAP data gives that
$\omega_0=-1.13^{+0.35}_{-0.26}$, $\omega_a=0.95^{+0.60}_{-1.95}$
and $\Omega_{m0}=0.28\pm 0.04$ with $\chi^2=175.62$. Substitute
the best fit parameters into Eq. (\ref{ztlind}), we get $z_{\rm
T}=0.56$. The results are summarized in Table \ref{fittab}.
The contour plots of $\omega_0$ and $\omega_a$ by fixing
$\Omega_{m0}$ at its best fit value $0.28$ are shown in Fig.
\ref{wqzl}. The evolution of $\omega(z)$ is shown in Fig.
\ref{obswqz}.

\begin{figure}
\centering
\includegraphics[width=12cm]{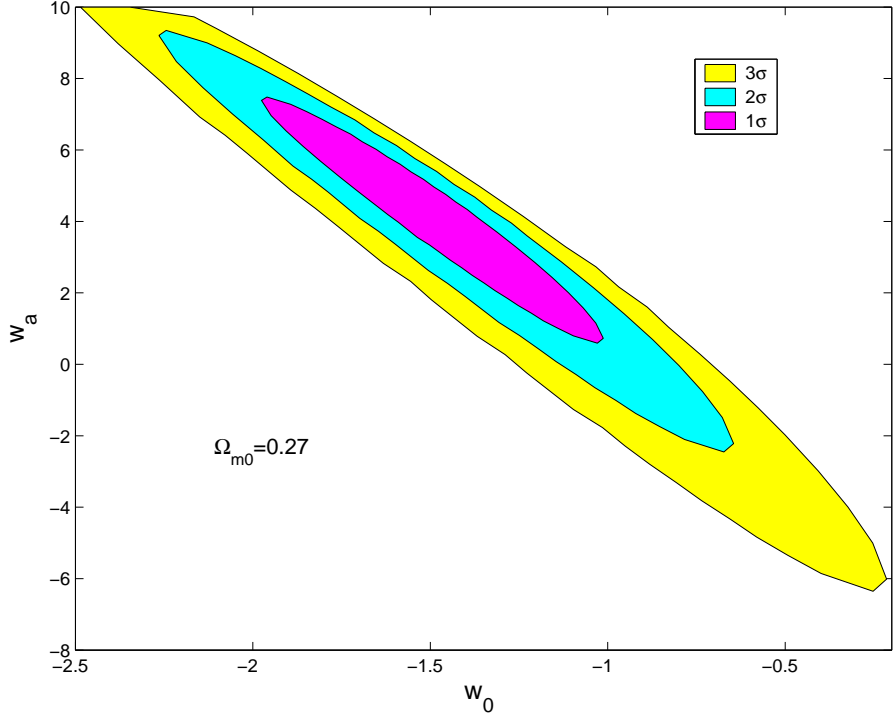}
\caption{The contour plot of $\omega_0$ and $\omega_a$ for the
parameterization $\omega=\omega_0+\omega_a z/(1+z)^2$}
\label{wqzl2}
\end{figure}

Model 4: Next we consider the following parameterization \cite{hkjbp},
\begin{equation}
\label{jbp}
\omega=\omega_0+{\omega_a
z\over (1+z)^2}.
\end{equation}
When $z\gg 1$, we have $\omega\sim \omega_0$. When $z\rightarrow
-1$, we have $\omega\rightarrow \pm \infty$. Substitute Eq.
(\ref{jbp}) into Eq. (\ref{conserv}), we get the dark energy
density
\begin{equation}
\label{jbpdark}
\Omega(z)=\Omega_0(1+z)^{3(1+\omega_0)}\exp(3\omega_az^2/2(1+z)^2).
\end{equation}
Substitute the above two equations (\ref{jbp}) and (\ref{jbpdark})
into Eq. (\ref{acceq}) and neglect the radiation contribution, we
find that $z_{\rm T}$ satisfies the following equation
\begin{equation}
\label{ztjbp}
\Omega_{m0}+(1-\Omega_{m0})\left(1+3\omega_0+{3\omega_a z\over
(1+z)^2}\right) (1+z)^{3\omega_0}\exp\left({3\omega_az^2\over
2(1+z)^2}\right)=0.
\end{equation}
The best fit to the combined SN Ia, SDSS and WMAP data gives
$\omega_0=-1.51\pm 0.6$, $\omega_a=4.3^{+3.8}_{-4.8}$
and $\Omega_{m0}=0.27\pm 0.04$ with $\chi^2=174.0$.
Substitute the best fit parameters into Eq. (\ref{ztjbp}), we get
$z_{\rm T}=0.38$. The results are summarized in Table \ref{fittab}.
The contour plots of $\omega_0$ and $\omega_a$ by
fixing $\Omega_{m0}$ at its best fit value $0.25$ are shown in Fig.
\ref{wqzl2}. The evolution of $\omega(z)$ is shown in Fig.
\ref{obswqz}.

\begin{table}
\caption{Summary of the best fit parameters.}
\label{fittab}
\begin{center}
\begin{tabular}{|c|c|c|c|c|c|c|}
  \hline
Model & $\Omega_{m0}$&$\Omega_{k0}$&$\omega_0$&$\omega_a$&$z_{\rm T}$&$\chi^2$\\
  \hline
1&$0.25\pm 0.05$&$-0.009\pm 0.050$&$-1.1\pm
0.2$&N/A&0.56&175.4\\\hline 2&$0.28\pm
0.04$&$-0.001^{+0.046}_{-0.045}$&$-0.97^{+0.17}_{-0.19}$&N/A&0.66&176.5\\\hline
3&$0.28\pm
0.04$&N/A&$-1.13^{+0.35}_{-0.26}$&$0.95^{+0.60}_{-1.95}$&0.56&175.62\\\hline
4&$0.27\pm 0.04$&N/A&$-1.51\pm 0.6$&$4.3^{+3.8}_{-4.8}$&0.38&174.0\\\hline
\end{tabular} \end{center} \end{table}

\section{Discussion}
We discussed two one-parameter dark energy parameterizations and
two two-parameter dark energy parameterizations. For the two
one-parameter dark energy parameterizations, we consider curved
cosmology, so that we have a total of three cosmological parameters: $\Omega_{m0}$, $\Omega_{k0}$
and $\omega_0$. For the two
two-parameter dark energy parameterizations, we consider flat
cosmology only, again there are three cosmological parameters: $\Omega_{m0}$, $\omega_0$ and $\omega_a$. These
different classes of three cosmological
parameter models fit the observational data almost equally well because they have almost
the same minimum value of $\chi^2$ as shown in table \ref{fittab}.
However, they have very different behavior. At early times,
the Universe is dominated by matter or radiation, the dark energy
is subdominant, so the contribution of dark energy to the background
evolution is not important and the data may not be used to
distinguish the early behavior of dark energy to the 0-th order.
From Fig. \ref{obswqz}, we see that the future behavior of
$\omega(z)$ are also very different. For the model 1,
$\omega(z)\rightarrow -\infty$ in the future. For the model
2, $\omega(z) \sim 0$ in the future. For the model
3, $\omega(z)\rightarrow -\infty$ in the future. For
the model 4, $\omega(z)\rightarrow +\infty$ in the
future. So the data may not used to distinguish the future
behavior of dark energy to the 0-th order either. We need to
invoke at least linear perturbation method to discuss dark energy
models. In \cite{pred}, the authors use the concept of the
minimal anti-trapped surface or the assumption that the energy
momentum content of the observable Universe does not change significantly
in comoving coordinates to study the fate of our universe. They found that
it is impossible to confirm the accelerating expansion with current observed
dark energy value $\Omega_{0}\sim 0.7$ if the dark energy is not a phantom. These
results more or less support our conclusion.
The dark energy in the models 1 and 2 tracked the matter in the past. The two models both
suggest that the Universe is almost spatially flat. All the models
suggest that $z_{\rm T}\sim 0.6$ and $\omega(0)\sim -1$. These
results are consistent with those derived from the simplest
$\Lambda$CDM model. However, it was found that $z_{\rm T}\sim
0.3$ by using SN Ia data only \cite{gong04}. So the results by
using combined SN Ia, SDSS and WMAP data are different from those
by using SN Ia data only. More thorough studies are needed to make
more concise conclusion. Finally, we would like to mention that the one and two-parameter representations
of dark energy models have their own limitations as discussed in \cite{bassett}. These
parameterizations do not accommodate the possibility of rapid evolution of dark energy.

\begin{acknowledgments}
Y. Gong is supported by CQUPT under grant
No. A2004-05, NNSFC under grant No. 10447008, CSTC under grant No.
2004BB8601 and SRF for ROCS, State Education Ministry. Y.Z.
Zhang's work was in part supported by NNSFC under Grant No.
90403032 and also by National Basic Research Program of China
under Grant No. 2003CB716300.
\end{acknowledgments}



\begin{thebibliography}{sp99}
\bibitem{sp99} S. Perlmutter {\it et al.},
Astrophy. J. {\bf 517}, 565 (1999).
\bibitem{gpm98} P.M. Garnavich  {\it et al.}, Astrophys. J. {\bf 493}, L53 (1998)
\bibitem{agr98} A.G. Riess  {\it et al.}, Astron. J. {\bf 116}, 1009 (1998).
\bibitem{review} V. Sahni  and A. A. Starobinsky, Int. J. Mod. Phys. D
{\bf 9}, 373 (2000).
\bibitem{padmanabhan03}T. Padmanabhan, Phys. Rep.
{\bf 380}, 235 (2003).
\bibitem{jwaa} J. Weller  and A. Albrecht, Phys. Rev. Lett. {\bf
86}, 1939 (2001).
\bibitem{hut} D. Huterer and M.S. Turner, Phys. Rev. D {\bf 64}, 123527 (2001).
\bibitem{jwaa1} J. Weller  and A. Albrecht, Phys. Rev. D {\bf 65}, 103512 (2002).
\bibitem{pastier} P. Astier, Phys. Lett. B {\bf 500}, 8 (2001).
\bibitem{polarski} M. Chevallier and  D. Polarski, Int. J. Mod.
Phys. D {\bf 10}, 213 (2001).
\bibitem{linder} E.V. Linder, Phys. Rev. Lett. {\bf 90}, 91301
(2003).
\bibitem{choudhury} T.R. Choudhury and  T. Padmanabhan, Astron. Astrophys. {\bf 429}, 807 (2005).
\bibitem{feng} B. Feng, X.L. Wang and X.M.
Zhang, Phys. Lett. B {\bf 607}, 35 (2005).
\bibitem{hkjbp}  H.K. Jassal, J.S. Bagla and T. Padmanabhan,
Mon. Not. Roy. Astron. Soc. {\bf 356}, L11 (2005).
\bibitem{gong04} Y. Gong, Class. Quantum Grav. {\bf 22}, 2121 (2005).
\bibitem{efstathiou} G. Efstathiou, Mon. Not. Roy. Soc. {\bf 310},
842 (1999)
\bibitem{gefstathiou} B.F. Gerke and G. Efstathiou, Mon. Not. Roy. Soc. {\bf
335}, 33 (2002).
\bibitem{pscejc} P.S. Corasaniti and E.J. Copeland, Phys. Rev. D {\bf
67}, 063521 (2003).
\bibitem{wetterich} C. Wetterich, Phys. Lett. B {\bf 594}, 17 (2004).
\bibitem{alam} U. Alam, V. Sahni, T.D. Saini and A.A.
Starobinsky, Mon. Not. Roy. Astron. Soc. {\bf 354}, 275 (2004).
\bibitem{alam1} U. Alam, V. Sahni and A.A.
Starobinsky, J. Cosmol. Astropart. Phys. JCAP {\bf 0406} (2004) 008.
\bibitem{daly} R.A. Daly and S.G. Djorgovski, Astrophys. J. {\bf 597}, 9
(2003).
\bibitem{daly1} R.A. Daly and S.G. Djorgovski, Astrophys. J. {\bf 612}, 652 (2004).
\bibitem{gong} Y. Gong, astro-ph/0401207, Int. J. Mod. Phys. D {\bf 14} (2005) 599.
\bibitem{jonsson} J. J\"{o}nsson, A. Goobar, R. Amanullah and L.
Bergstr\"{o}m, J. Cosmol. Astropart. Phys., JCAP {\bf 0409} (2004) 007.
\bibitem{wang} Y. Wang and P. Mukherjee,
Astrophys. J. {\bf 606}, 654 (2004).
\bibitem{wang1} Y. Wang and M. Tegmark, Phys.
Rev. Lett. {\bf 92}, 241302 (2004).
\bibitem{cardone} V.F. Cardone, A. Troisi and S. Capozziello,
Phys. Rev. D {\bf 69}, 083517 (2004).
\bibitem{huterer} D. Huterer and A. Cooray, Phys. Rev. D {\bf 71}, 023506 (2005).
\bibitem{riess} A.G. Riess {\it et al.}, Astrophys. J. {\bf 607}, 665 (2004).
\bibitem{sdss} D.J. Eisenstein {\it et al.}, astro-ph/0501171.
\bibitem{wmap} C.L. Bennett {\it et al.}, Astrophys. J. Supp. Ser. {\bf 148}, 1
(2003).
\bibitem{sandvik} H.B. Sandvik, M. Tegmark, M. Zaldarriaga and I.
Waga, Phys. Rev. D {\bf 69}, 123524 (2004).
\bibitem{bento04} M.C. Bento, O. Bertolami and A.A. Sen, Phys.
Rev. D {\bf 70}, 083519 (2004).
\bibitem{rrries} R.R.R. Reis, M. Makler and I. Waga,
Class. Quantum Grav. {\bf 22}, 353 (2005).
\bibitem{guo} Z.K. Guo, R.G. Cai and Y.Z. Zhang, astro-ph/0412624,
J. Cosmol. Astropart. Phys., JCAP {\bf 0505} (2005) 002.
\bibitem{pred} T. Vachaspati and M. Trodden, Phys. Rev. D {\bf 61}, 023502 (2000);
G. Starkman, M. Trodden and T. Vachaspati, Phys. Rev. Lett. {\bf 83}, 1510 (1999);
P.P. Avelino, J.P.M. de Carvalho and C.J.A.P. Martins, Phys. Lett. B {\bf 501}, 257 (2001);
D. Huterer, G.D. Starkman and M. Trodden, Phys. Rev. D {\bf 66}, 043511 (2002).
\bibitem{bassett} B.A. Bassett, P.S. Corasaniti and M. Kunz, Astrophys. J. {\bf 617}, L1 (2004).

\end{thebibliography}
\end{document}